\RequirePackage{fix-cm}
\documentclass[twocolumn]{svjour3}          
\smartqed  
\usepackage{amssymb}
\usepackage{amsmath}
\usepackage{graphicx}
\usepackage{amsmath}
\usepackage{color}
\usepackage[version=3]{mhchem}
\usepackage{tabularx}
\usepackage{textcomp}
\begin{document}

\title{Superconductivity in the noncentrosymetric \ce{Li_2Pd_{3-x}Ag_xB} with x=0.0, 0.1 and 0.3}



\author{A.A. Castro \and O. Olic\'on \and F. Morales \and R. Escudero}


\institute{A.A. Castro \and O. Olic\'on \and F. Morales \and R. Escudero \at
              Instituto de Investigaciones en Materiales, Universidad Nacional
              Aut\'onoma de M\'exico, Ciudad de M\'exico, 04510, M\'exico.\\
              \email{aacastroespinosa@gmail.com}\\
           }

\date{Received: date / Accepted: date}

\maketitle 

\begin{abstract}
We studied the partial substitution of Ag in the noncentrosymetric superconducting compounds Li$_2$Pd$_{3-x}$Ag$_x$B with x=0.0, 0.1 and 0.3. Magnetization, resistivity and specific heat measurements were performed in this system.  From electrical measurements we obtained the   critical temperature $T_c$. The upper critical field  at zero temperature was determined with Werthamer-Helfand equation, $\mu_0H^{WHH}_{c2}(0)$, and $\mu_0H^{linear}_{c2}(0)$ and the coherence length, $\xi(0)$. The Value of  $\mu_0H^{linear}_{c2}(0)$ and $\mu_0H^{WHH}_{c2}(0)$ is lower than  the calculated  paramagnetic Pauli limit suggesting  that the  spin-triplet pairing is weak in the system.  From specific heat measurements we obtained the parameters of the normal state;  the Sommerfeld coefficient, $\gamma$, the  electronic density of states at the Fermi level, $N(E_F)$ and the Debye temperature, $\theta_D$. In the superconducting state the obtained values were;  $\Delta C/\gamma T_c$, the electron phonon coupling, $\lambda_{e-ph}$, the superconducting  energy gap, $2\Delta_0$, and the ratio $2\Delta_0 /k_BT_c$. The electronic specific heat  is well described by the BCS  theory, suggesting that the energy gap is isotropic  and in a strong coupling state.  


\end{abstract}

\section{Introduction}
\label{intro}
Since the discovery of superconductivity in 1911,  the search for novel superconductors specially materials with higher critical temperatures have been carried out intensively. Superconducting materials can be classified in many  categories: metal-based system, copper-oxygen,  and iron pnictides/chalcogenides (iron-based superconductors), and others \cite{Hosono}. 

For instances,  Iron-based superconductors are ideal candidates for several applications of superconductivity:  wires, tapes, and coated conductors for high magnetic fields. The pollycrystalline compound FeSe, is an interesting material for bulk applications like superconducting trapped-field magnets or super-magnets \cite{Liu2015,Hsu,Koblischka}.  

Superconductivity in materials without inversion symmetry   exhibit engaging properties because a strong modification of the  electronic band structure  caused by antisymmetric spin-orbit coupling ASOC \cite{bauer}. It has been shown that ASOC acts unfavorable on spin-triplet pairing states \cite{anderson}. For spin single states the influence of ASOC is minor. The presence of ASOC leads to a splitting of the  electronic bands by lifting the spin degeneracy and consequently  splitting the Fermi surface. In noncentrosymetric superconductors the two electronic bands  forming Cooper pairs belong to two different Fermi surface corresponding to the spin-up and spin-down bands \cite{Anand}. The  presence, or  absence of inversion symmetry may leads to a mixture of spin-singlet and spin-triplet coupling \cite{Anand,gorkov,Frigeri,Sigrist,Smidman}. Depending on the type of the pairing interactions, the gap function may be  characterized by the presence of line nodes \cite{Frigeri}.

Noncentrosymmetric heavy fermion superconductors such as \ce{CeRhSi_3} \cite{kimura}, \ce{CeIrSi_3} \cite{Sugitani}, \ce{UIr} \cite{Akazawa}, \ce{CePt_3Si} \cite{Yogi,Izawa}  have attracted attention because  the unconventional behavior in these strongly correlated electronic  compounds. On  the other side, transition-metal compounds like \ce{Li_2Pd_3B} \cite{togano}, Li$_2$Pd$_{3-x}$Cu$_x$B with x=0.0, 0.1 and 0.3 \cite{Castro}, \ce{Li_2Pt_3B} \cite{Badica}, \ce{Mg_{10}Ir_{19}B_{16}} \cite{Klimczuk}, \ce{Rh_2Ga_9}, \ce{Ir_2Ga_9} \cite{Shibayama}, \ce{Nb_{0.18}Re_{0.82}} \cite{Karki} are more suitable for exploring the issue of inversion symmetry breaking.
In previous reports, the ternary metallic borides \ce{Li_2Pd_3B} were   classified as  noncentrosymetric conventional superconductor without strong electronic correlation \cite{togano,mani,Nishiyama}.

One antecedent different to this issue is that  the  well known 
ternary metallic boride \ce{Li_2Pd_3B}, discovered by Togano with transition temperature about 8 K \cite{togano}, was classified as  noncentrosymetric conventional superconductor without strong electronic correlation \cite{mani,Nishiyama}.

In this study we present  the effect of substitution of Pd with a nonmagnetic element, in an early reported \cite{Castro}. We choose the substitution of Pd with Ag, a non magnetic element. The reason was  because is isostructural and the atomic ratio of Ag is bigger  than  Pd. The difference in the atomic ratio should generates a internal  pressure effect in the unit cell and  modify the superconducting properties. In this work  we study the characteristics of this compound: crystalline structure,  magnetic and electronic  specific heat \ce{Li_2Pd_{3-x}Ag_xB} with three Ag  compositions; x=0.0, 0.1 and 0.3.  We compare and observe the Pauli limit with the upper critical field  because is a criterion  related to the importance of the spin-triplet piring component. We estimate the parameters of the normal and superconducting state, and the strength of the electron-phonon  coupling.

\section{Experimental details}
Polycrystalline samples of \ce{Li_2Pd_{3-x}Ag_xB} with three compositions; x=0.0, 0.1 and 0.3 were synthesized by  two steps in an arc-melting apparatus in order
to minimize  losses of Li by evaporation. Stoichiometric samples were prepared using 20\% of Li (99 +\%) excess, Pd (99.95\%), Ag (99.95\%) and
B (99.99\%) powders as precursors. The melting was performed  in a chamber with  high pure argon atmosphere.
X ray diffraction patterns were measured at room
temperature in a Siemens (D5000) diffractometer with Co $K_{\alpha}$ radiation ($\lambda$=1.79026 \AA)
and Fe filter in steps of .015$^{\circ}$ at 8 s in the 2$\theta$ range of 20$^{\circ}$-110$^{\circ}$.  Rietveld analysis of the diffraction patterns were performed with  MAUD program \cite{maud}.

Electrical resistance versus temperature $R(T)$ and magnetic field were measured  in a Physical Properties Measurement System, PPMS (Quantum Design). $R(T)$  without magnetic field were measured from 300 K to 2 K. The magnetoresistance measurements were determined between 2 K and 10 K with applied magnetic field  between  0 and 40 kOe.
Magnetization measurements were determined in a MPMS  magnetometer (Quantum Design). Temperature
dependence of the magnetic susceptibilities were investigated under zero field cooling (ZFC) and field cooling (FC) modes.

Specific heat measurements were determined using a relaxation method between room temperature and 2 K in the  PPMS system. The sample was attached to
the measuring stage using Apiezon N grease to ensure a good thermal contact. The contribution to the heat capacity of the sample holder and grease was  subtracted from the sample measurements.

\section{Results and discussion}

\subsection{Structural characterization}

\begin{figure}[ht]
\begin{center}
\includegraphics[scale=0.3]{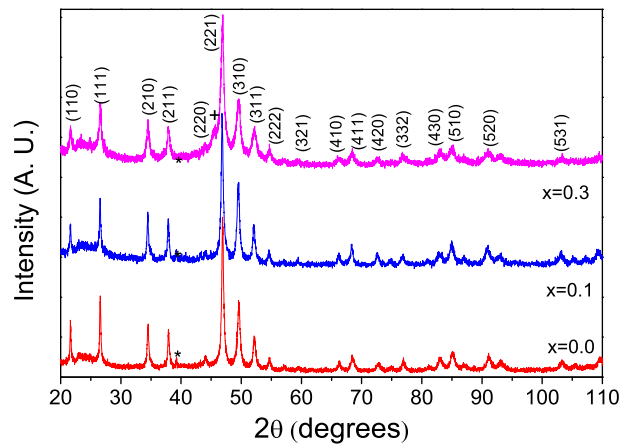}
\caption{\label{RxAG} X-ray diffraction patterns for \ce{Li_2Pd_{3-x}Ag_xB} with x=0, 0.1 and 0.3. The reflections observed at 39.3$^{\circ}$ correspond
to a tiny impurity of \ce{Pd_2B_5} and \ce{PdB_2} respectively.}
\end{center}
\end{figure}

X-ray diffraction patterns for the polycrystalline samples  \ce{Li_2Pd_{3-x}Ag_xB} with the three compositions: x=0.0, 0.1 and 0.3 are shown in Figure \ref{RxAG}. The main features of the patterns correspond to a  cubic structure $\mathrm{P4_{3}32}$. A small amount of impurities \ce{Pd_2B_5} ($^{\ast}$) and \ce{PdB_2}
($^{+}$) were detected. It is worth mentioning that \ce{PdB_2} is non-superconducting \cite{Chen}.

\begin{figure}[ht]
\begin{center}
\includegraphics[scale=0.3]{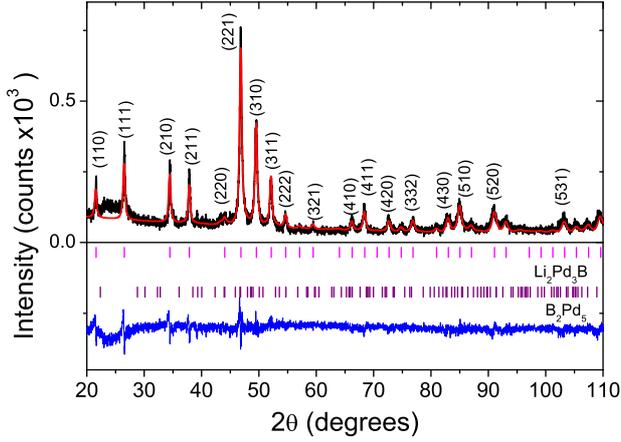}
\caption{\label{RefAg01} Rietveld results for the \ce{Li_2Pd_{2.8}Ag_{0.1}B} sample along with experimental (black line) and calculated pattern
(red line). Vertical lines are the reflections of \ce{Li_2Pd_{3}B} and \ce{Pd_2B_5}. The line at the bottom is the difference between experimental and refined pattern}
\end{center}
\end{figure}

The X-Ray diffraction patterns were Rietveld-fitted considering the possibility that Ag ions can occupy Pd sites. Figure \ref{RefAg01} shows the refined pattern of \ce{Li_2Pd_{2.9}Ag_{0.1}B} sample. Vertical lines, at the bottom of the panel are the reported reflections (ICSD 84931),  the Miller indexes of each plane are indicated.

The structural parameters and R-factors are summarized in  Table \ref{table:parametrosAg}. The structural parameters and R-factors obtained from the Rietveld analysis, for the three samples in study, shows increase of the lattice parameter, $a$, as the nominal concentration of Ag increases.  This results from the atomic radius difference between Pd (1.28\AA{}) and Ag (1.34\AA{}). This certainly suggests that Ag has substituted Pd in the \ce{Li_2Pd_{3-x}Ag_xB} with x=0.0, 0.1 and 0.3 system. At concentrations x$>$0.3 traces of Ag were detected, so we considered just x$=0.0,0.1$ and 0.3.

\begin{table}[h]
 \caption{Structural parameters obtained from the Rietveld fitting of the X-ray diffraction patterns of the studied  compound with  {Ag} at 295 K.}
 \resizebox{8.4cm}{!}{
  \begin{tabular}{c c c c}
 \hline
  &x=0.0 & x=0.1 &x=0.3 \\
 \hline
  a (\AA{})                     & 6.7427(3) & 6.7548(3)  & 6.7637(9)\\
 V (\AA{}$^3$)                      & 306.6     & 308.2  & 309.4    \\
 $R_{w}$(\%)                    & 15.82     & 16.28      & 14.34    \\
 $R_{b}$(\%)                    & 12.28     & 12.72      & 11.08    \\
 $R_{exp}$(\%)                  & 10.87     & 11.98      & 10.37    \\
 $\chi^{2}$(\%)                 & 1.45      & 1.35       & 1.38     \\
 $\mathrm{Li_{2}Pd_{3-x}Ag_xB}$ (\%)  & 90.5590   & 97.5043    & 80.1257\\
 $\mathrm{B_{2}Pd_{5}}$ (\%)    & 9.4490    & 2.4956     & 19.0506  \\
 $\mathrm{Pd_{2}B}$ (\%)        & -         & -          & 0.8235   \\
 \hline
 \end{tabular}

 \label{table:parametrosAg}}
\end{table}
\subsection{Magnetic and electrical measurements}
\begin{figure}[h]
\begin{center}
\includegraphics[scale=0.3]{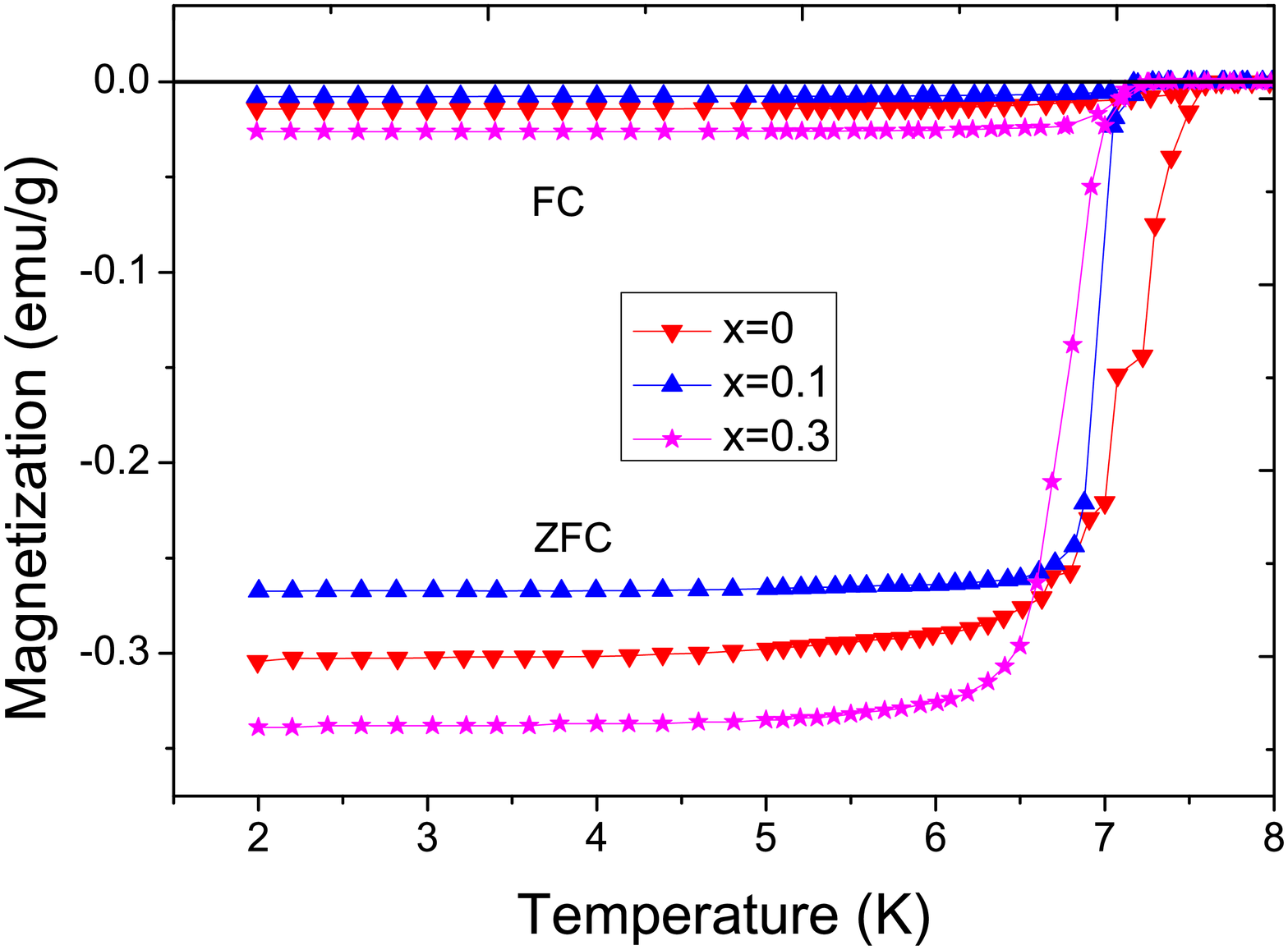}
\caption{\label{MT} Magnetization as a function of temperature of \ce{Li_{2}Pd_{3-x}Ag_{x}B} at constant magnetic field of 20 Oe. }
\end{center}
\end{figure}

Magnetization curves in zero-field-cooling (ZFC) and field-cooling (FC) modes  as  function of temperature under 20 Oe applied magnetic field are shown in Figure \ref{MT}. A drop of magnetization below 8 K was observed for
the samples in this work. The critical temperature, $T_c$, was determined as the onset  of the transition. The superconducting transition decreases gradually with  doping of Ag. The samples in studied present a low fraction corresponding to the Meissner efect (FC) comparable to the shielding effect (FC) which is common to see in bulk pollycrystalline metallic superconductors \cite{togano,Badica}. 

\begin{figure}[h]
\begin{center}
\includegraphics[scale=0.3]{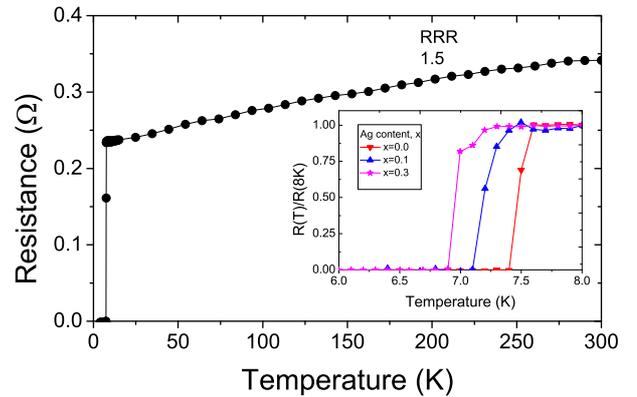}
\caption{\label{RvsT} Resistance vs Temperature  of \ce{Li_2Pd_3B}, the residual resistance ratio (RRR) of this sample is 1.5. The inset shows the normalized resistance $R/R(8K)$ curves above 2 K and up to 8 K, close to the superconductor transition, of the measured samples}.
\end{center}
\end{figure}

The electrical resistance as a function of temperature $R(T)$ from 2 to 300 K of the parent compound, \ce{Li_2Pd_3B} is shown in Figure \ref{RvsT}. The resistance
shows a  small decreasing  behavior below room temperature. The residual resistance ratio $(RRR=R_{300K}/R_{8K})$
is indicated. The $RRR$ values provide a qualitative information about electron scattering by impurities and vacancies. For the samples studied  the $RRR$ values were  1.5, 3.0 and 2.9  for the three compositions (x=0.0, 0.1 and 0.3) those values are in the range of 1.4 to 6.5  of the  reported values for \ce{Li_2Pd_3B}  \cite{mani,togano}. The small value of RRR obtained in this study could suggests that  the normal-state
resistivity has a  weak temperature dependence on the electrical transport and  very sensitive to impurities and disorder.

The inset of figure \ref{RvsT} shows the normalized resistance with resistance value at 8 K. The superconducting transitions are sharp with a transition width of 0.21 K for the samples under study.  The critical temperature was determined in the point at which the first derivative of the  R(T) curve reaches its maximum value. The $T_c$ values are 7.51 K, 7.19 K and 6.99 K for samples with
x=0.0, 0.1 and 0.3 respectively. It can be seen that $T_c$ decreases as the Ag content increases.

\begin{figure}[ht]
\begin{center}
\includegraphics[scale=0.34]{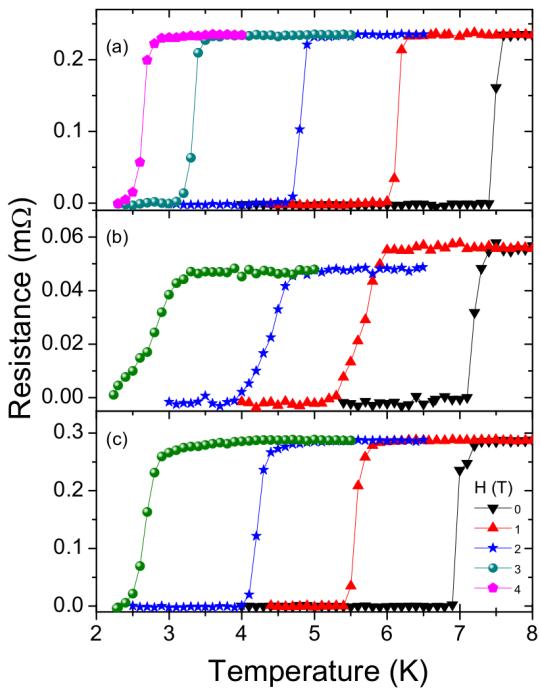}
\caption{\label{MRT} Magnetoresistance measurements with temperature and magnetic field for \ce{Li_2Pd_{3-x}Ag_xB} (a) x=0.0, (b) x=0.1 and (c) x=0.3. The symbols indicate the applied magnetic fields for the three samples.}
\end{center}
\end{figure}

\begin{figure}[ht]
\begin{center}
\includegraphics[scale=0.33]{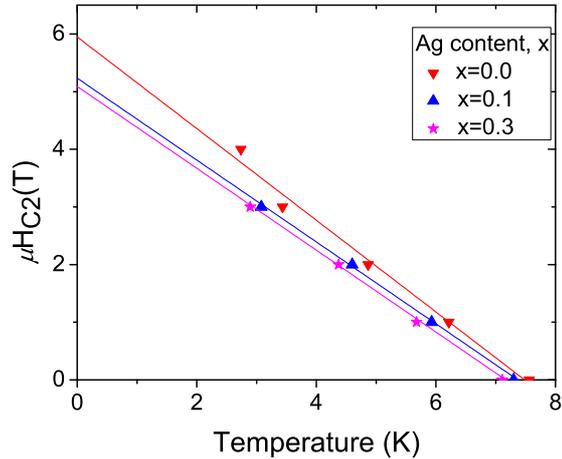}
\caption{\label{HC2} Upper critical field as a function of temperature determined from magnetoresistance measurements of \ce{Li_2Pd_{3-x}Ag_xB} (x=0.0, 0.1 and 0.3). The continuous lines are linear fit of data and its extrapolation to $T=0$ K gives the values of $H_{c2}^{linear}(0)$}. 
\end{center}
\end{figure}

The resistance as  function of temperature and magnetic field for the three studied samples in the 2-8 K temperature range under 0-4 T magnetic field is shown in Figure \ref{MRT}. The transition to the superconducting state is shifting to lower temperatures under the increase of DC magnetic field. From this curves we extracted  the variation of the $T_c$ with the applied field shown in Figure \ref{HC2}. The  extrapolation of the linear fit of data the upper critical field at $T=0$ K, $H_{c2}^{linear}(0)$, with values of 6.0$\pm0.3$, 5.2$\pm0.1$ and 5.1$\pm0.1$ for the three  samples.

The upper critical field at $0$ K was estimated using the Werthamer-Halfand-Hohenberg theory \cite{WHH}:

\begin{equation}
 H_{c2}^{WHH}(0)=0.693 T_c \left(\frac{dH_{c2}}{dT}\right)_{T=T_c},
\end{equation}
this theory has been employed in several studies of noncentrosymmetric superconducting systems \cite{Castro,Badica,Klimczuk,Karki,mani,Biswas}.

The values of  $\mu_{0}H_{c2}^{WHH}(0)$ obtained are 4.95 T, 4.43 T and 4.29 T for x=0.0, 0.1 and 0.3 respectively. The value of $\mu_{0}H_{c2}^{WHH}(0)$ (4.95 T) for the sample \ce{Li_2Pd_3B} is comparable with the reported value of 4.8 T \cite{togano}.
From the $\mu_{0}H_{c2}^{WHH}(0)$ values we calculate the
coherence length, $\xi(0)$, using the Ginzburg-Landau equation:

\begin{equation}
 H_{c2}(0)=\frac{\Phi_0}{2\pi\xi^2_0},
\end{equation}
where  $\Phi_0=2.0678\times 10^{-15}$ T m$^2$ is the magnetic flux quantum. The values of $\xi_0$ are 8.17, 8.64 and 8.78 nm for x=0.0, 0.1 and 0.3, respectively.

The obtained Pauli limit field, $\mu_0H^{Pauli}=\Delta_0/\mu_B \surd   2$ \cite{Clogston},
takes values of 17.83 T (x=0.0), 16.91 T (x=0.1) and 17.77 T (x=0.3). It is noteworthy, that these values are higher that the values obtained  from the upper critical fields. Accordingly,  if the upper critical field in a  single crystal exceeds the Pauli limit   this suggests  and implies a substantial contribution from the  spin-triplet component to the pairing amplitude, presumably because  broken inversion symmetry \cite{Karki}. However, in this  study the Pauli  limiting field   is greater than the upper critical fields $\mu_{0}H_{c2}(0)^{WHH}$ and  $\mu_{0}H_{c2}(0)^{linear}$, suggesting that the Cooper pairs are in spin-singlet state. The upper critical fields  values obtained  from a  linear extrapolation and the WHH theory shows a decrease while increasing Ag substitution: As a result a increase in $\xi_{0}$ values is obtained for the samples in study.

\subsection{Specific heat}
In the normal state, above the transition temperature, the specific heat data is well fitted by a sum of the electronic and lattice contribution \cite{tari}:
\begin{equation}
 C_p=\gamma T+\beta T^3.
\end{equation}

\begin{figure}[ht]
\begin{center}
\includegraphics[scale=0.26]{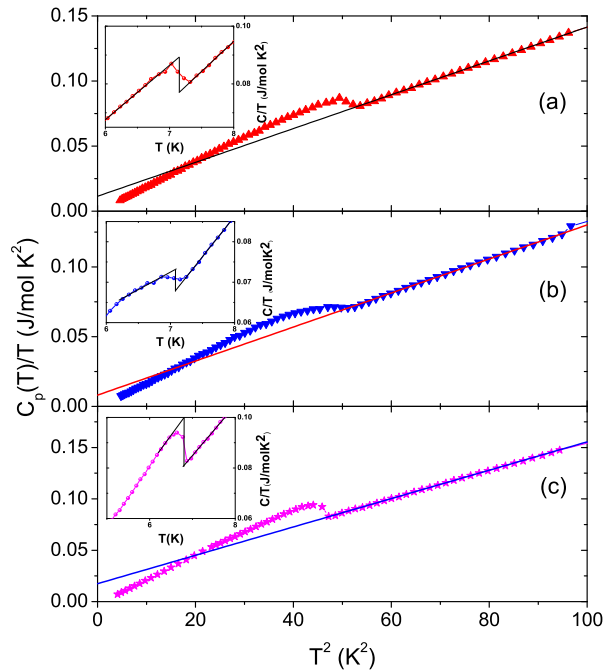}
\caption{\label{heat1} $C_p/T$ versus $T^2$ curves of \ce{Li_2Pd_{3-x}Ag_xB}; (a) x=0.0, (b) x=0.1 and (c) x=0.3. The solid line represents a linear fit to the specific heat in the normal state. The inset shows $C_p/T$ versus $T$ to determine the specific heat jump at the superconducting transition.}
\end{center}
\end{figure}

A linear fit to $C_p/T$ vs $T^2$ plot of the \ce{Li_2Pd_{3-x}Ag_xB} with x=0.0, 0.1 and 0.3 are shown in Figure \ref{heat1} yielding  the Sommerfeld coefficient
$\gamma$ and the Debye constant $\beta$. For the sample without Ag the value of $\gamma$ is 1.26 $\mathrm{mJ/mol K^2}$, this value is a little bit higher than the reported previously, which varies from 8.3 to 9.8 $\mathrm{mJ/mol K^2}$ \cite{Takeya2005,Takeya2}. From the value of $\beta$ we estimate the Debye temperature using the relation  
\begin{equation}
 \theta_D=\left(\frac{12n\pi^4R}{5\beta}\right)^{\frac{1}{3}}\end{equation}
where  $n$ is the number of atoms per formula unit. The value of $\theta_D$ for the \ce{Li_2Pd_3B} sample is 207.8 K, which agrees with the reported values of 221 K \cite{Takeya2005} and 202 K \cite{Takeya2007111}. On the other hand using the experimental Sommerfeld constant $\gamma$, the  electronic density of states at Fermi level $N(E_F)$ can be deduced using the formula:
\begin{equation}
 N(E_F)=\frac{3\gamma}{\pi^2k_B^2N_A}.
\end{equation}
$k_B$ is the  Boltzmann's constant and $N_A$ the  Avogadro's number. The parameters obtained are shown in Table \ref{table:heatAg}. It can be seen that there is a non-monotonic  variation of $N(E_F)$ and $\gamma$ as the concentration of Ag is increased. This may be due to the impurity of \ce{B_2Pd_5} on the sample \ce{Li_2Pd_{2.7}Ag_{0.3}B}.

In the inset of Figure \ref{heat1}, $C_p/T$ vs $T$ is plotted  detail of the specific heat jump at the thermodynamic transition. 

From the critical temperature $T_c$ and Debye
temperature  $\theta_D$, we can estimate the electron-phonon coupling constant $\lambda_{e-ph}$ with the  Mc Millan's relation \cite{McMillan} given below
\begin{equation}
 \lambda_{e-ph}=\frac{1.04+\mu^{*}\ln(\theta_D / 1.45 T_c)}{(1-0.62\mu^*)\ln(\theta_D/1.45T_c)-1.04}
\end{equation}
where $\mu^*$ represents the screened repulsive Coulomb potential,  usually in the range 0.1-0.15, in this work we used the value of 0.13. This value is normally used for intermetallic superconductors \cite{Takeya2005,McMillan}. The values of $\lambda_{e-ph}$ obtained in this study range from 0.84 and 0.81 for different contents of Ag. The value reported for the \ce{Li_2Pd_3B} compound is $\lambda_{e-ph}=1.09$ \cite{Takeya2005}, which is higher that the obtained in this work. The values of $\lambda_{e-ph}$ are shown in Table \ref{table:heatAg} and allows to classify the system \ce{Li_2Pd_{3-x}Ag_xB} with x=0.0, 0.1 and 0.3 in a moderated coupled superconductor state.

\begin{figure}
\begin{center}
\includegraphics[scale=0.3]{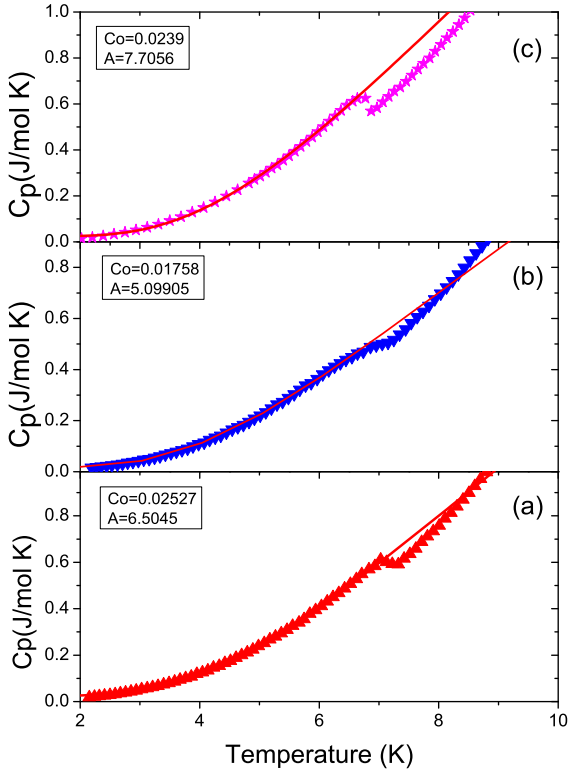}
\caption{\label{BRECHAS}  $C_p$ versus $T$ curves of \ce{Li_2Pd_{3-x}Ag_xB}; (a) x=0.0, (b) x=0.1 and (c) x=0.3. The solid line represents a linear fit to equation \ref{cp}.}
\end{center}
\end{figure}

At low temperatures the specific heat fit well to an exponential decay \cite{Escudero}:
\begin{equation}
 C_p=C_0+A\exp\left(\frac{-\Delta_0}{k_BT}\right),
 \label{cp}
\end{equation}
where $C_0$ is assumed as background of $C_p$. In order to determine the superconducting  energy gap, $2\Delta_0$,  the $C_p(T)$ data were fitted with equation \ref{cp}. The fit is shown as a continuous line  in Figure \ref{BRECHAS}. The  obtained  values of the superconducting gap, $2\Delta_0$ and $2\Delta_0/k_B T_{C}$ for the samples under study are shown in Table \ref{table:heatAg}. 

Several values of the superconducting energy gap of \ce{Li_2Pd_3B} have been reported, from 2.55 to 3.2 meV \cite{Takeya2005,Takeya2,Takeya20081078}, the value of $2\Delta_0=2.92$ meV, obtained in this work, agrees with these values. The values of $2\Delta_0/k_BT_c$ obtained in this work increase from 4.60 for x=0.0 to 5.03 for x=0.3, these values are  higher than  the value predicted by BCS, $2\Delta_0/k_BT_c=3.52$,  and to the reported value of 3.94 \cite{Takeya2005}. Our results indicate that the system \ce{Li_2Pd_{3-x}Ag_xB} with x=0.0,
0.1 and 0.3 is a superconducting system  with strong electron-phonon coupling.

\begin{table}[h]
 \caption{Parameters obtained from the specific heat of the system $\mathrm{Li_{2}Pd_{3-x}Ag_{x}B}$ (x=0.0, 0.1 and 0.3);  Sommerfeld coefficient $\gamma$, density of states at Fermi level $N(E_F)$, Debye constant $\beta$,  Debye temperature $\theta_{D}$, transition temperature $\mathrm{T_{C}}$, specific heat jump at $T_C$ divided by $\gamma T_{C}$
 $\Delta C/\gamma T_{C}$, electron$-$fonon coupling constant $\mathrm{\lambda_{e-ph}}$, superconducting energy gap
 2$\mathrm{\Delta(0)}$ and the ratio $2\Delta/k_B T_C$ .}
  \resizebox{8.4cm}{!}{
 \begin{tabular}{l c c c  }
 \hline
x                         & 0.0                  & 0.1                 & 0.3                   \\
\hline
$\gamma$ (mJ/mol K$^2$)       &11.26$\pm$0.31        &8.11$\pm$0.37        &17.48$\pm$0.26  \\
$N(E_F)$ (eV$^{-1}$)      & 2.43$\pm$0.065       &1.72$\pm$0.078       &3.70$\pm$0.055  \\
$\beta$ (mJ/mol K$^4$)    & 1.3$\pm$0.004        &1.22$\pm$0.005       &1.38$\pm$0.004  \\
$\theta_D$ (K)            &208                   &212                  &204              \\
\hline
$T_C$ (K)                 & 7.36                 &6.89                 &6.72            \\
$\Delta C /\gamma T_{C}$  & 0.96$\pm$0.03        &1.14$\pm$0.05        &0.89$\pm$0.01   \\
$\lambda_{e-f}$           & 0.84                 &0.81                 &0.82            \\
$2\Delta(0)$ (meV)        &2.92$\pm$0.03         &2.77$\pm$0.02        &2.91$\pm$0.04   \\
$2\Delta(0)/k_B T_{C}$    &4.60$\pm$ 0.03        &4.67$\pm$ 0.03       &5.03$\pm$0.07    \\
\hline
 \end{tabular}
 \label{table:heatAg}}
\end{table}

 The specific heat jump at $T_c$, $\Delta C/\gamma T_c$ determined from $C_P(T)$ data are lower than the BCS predicted value, 1.43, and to the reported values in \ce{Li_2Pd_3B} \cite{Takeya2005,Takeya2,Takeya20081078} that varies from 1.6 to 2. Differences of $\Delta C/\gamma T_c$ between different reports may be due to the difficulties in achieving the stoichiometry of Li and disorder in the samples.
 
 We compare the results obtained in this work with the other  published \ce{Li_2Pd_{3-x}Cu_xB} with x=0.0, 0.1 and 0.2, where the lattice parameter decreases, $a$, as the nominal concentration of Cu increases. It is clear that the substitution of Pd with Ag and Cu generates an internal pressure in the unit cell and therefore a decrement of $T_c$, which can be related with a decrease in the $2\Delta(0)$.   

\section{Conclusions}
In this work we  reported the synthesis of three polycrystalline samples of \ce{Li_2Pd_{3-x}Ag_xB} with compositions:  x=0.0, 0.1 and 0.3 prepared by arc melting technique. Rietveld analysis of the  X-ray diffraction patterns shows an increase in  lattice parameters  as  Ag content increases.
Magnetization, electrical transport and specific heat measurements confirms that the  superconducting transition temperature, $T_c$, decreases as the content of  Ag is increased.
The upper critical fields $\mu_{0}H_{c2}^{WHH}(0)$ and $\mu_{0}H_{c2}^{linear}(0)$ are lower than the Pauli limiting field which suggests that the  pairng is in a spin-singlet state. The behavior of the low-temperature specific heat, allows  to classify this  system as  a superconductors with an isotropic energy gap as a conventional BCS superconductor, where the ratio, $2\Delta_0/k_B T_c$, values are indicative  of   strong coupling, contrary to the determined in other studies already published. We noted than Ag sustitution decreases the electron $N(E_F)$, observed in the decreasing of the energy gap. For the two compounds with Ag the evidence is not so clear because  we found small impurities. However  in the most pure sample the effect is quite clear.

\section*{Acknowledgments}
We acknowledge to DGAPA-UNAM IT100217,  M. C,  A. Bobadilla for helium supply, and  also  to A. Lopez, Alan Dierick Ortega Guti\'errez and A. Pompa-Garcia for help in computing details. A. A. Castro thanks CONACYT, for the support through the Posdoctoral Scholarship and C. Gonz\'alez for help in computing details.

\section*{Conflict of interest}
The authors declare that they have not conflict of interest.

\bibliographystyle{spphys}      
\bibliography{references}

\end{document}